# Maximum likelihood fitting of acyclic directed mixed graphs to binary data

**Robin J. Evans**
Department of Statistics
University of Washington
rje42@stat.washington.edu

**Thomas S. Richardson**
Department of Statistics
University of Washington
tsr@stat.washington.edu

## Abstract

Acyclic directed mixed graphs, also known as semi-Markov models represent the conditional independence structure induced on an observed margin by a DAG model with latent variables. In this paper we present the first method for fitting these models to binary data using maximum likelihood estimation.

## 1 Introduction

Acyclic directed mixed graphs (ADMGs), also known as semi-Markov models, contain directed ($\rightarrow$) and bi-directed ($\leftrightarrow$) edges subject to the restriction that there are no directed cycles. Such graphs are useful for representing the independence structure arising from a directed acyclic graph (DAG) model with hidden variables. More recently they have proved useful in characterizing the set of intervention distributions that are identifiable (Huang and Valtorta, 2006; Shpitser and Pearl, 2006). Ancestral graphs without undirected edges (Richardson and Spirtes, 2002) are a subclass of ADMGs.

The associated statistical models have many desirable properties: they are everywhere identified and they form curved exponential families with a well-defined dimension.

In contrast, latent variable (LV) models lack many of these properties. In particular: LV models are not always identified; inferences may be sensitive to assumptions made about the state spaces for the unobserved variables; LV models may contain singularities, at which the asymptotics are irregular (Drton, 2009); LV models do not form a tractable search space: an arbitrary number of hidden variables may be incorporated, so the class contains infinitely many different structures relating a finite set of variables.

Other approaches, such as probit models (Silva and

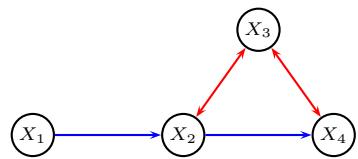

Figure 1: A mixed graph $\mathcal{G}_1$. (Color is used to distinguish the two edge types.)

Ghahramani, 2009) or cumulative distribution networks (Huang and Frey, 2008) provide a more parsimonious sub-class of models with ADMG Markov structure, but impose additional non-Markovian constraints.

The global Markov property for ADMGs is given by the natural extension of d-separation to graphs with bi-directed edges. Richardson (2003) provided a local Markov property for acyclic directed mixed graphs that was equivalent to the global property for all distributions. Richardson (2009) introduced a general factorization criterion for these models, and a parametrization in the binary case.

Mixed graphs allow a more complex range of hypotheses to be included. We use the notation of Dawid (1979): for random variables $X, Y$ and $Z$, we denote the statement '$X$ is independent of $Y$ conditional on $Z$' by $X \perp\!\!\!\perp Y | Z$; if $Z$ is trivial we write $X \perp\!\!\!\perp Y$.

The graph in figure 1, for example, encodes the independence relations

$$X_1 \perp\!\!\!\perp X_4 | X_2 \qquad X_1 \perp\!\!\!\perp X_3$$

but not $X_1 \perp\!\!\!\perp X_4 | (X_2, X_3)$ or any others in general.

In this paper we first give a result which allows some simplification of the parametrization, and build on this to provide the first method for fitting ADMGs to multivariate binary data. Since the graph corresponds to a curved exponential family and our fitting method uses maximum likelihood techniques, we can perform model selection with BIC or $\chi^2$ goodness of fit tests.

In section 2 we introduce our notation, and key concepts for ADMGs; in section 3 we discuss a way of representing the parametrization of ADMGs, and in section 4 we use this in our fitting algorithm. Examples are given in section 5, with a brief discussion of computation time in section 6. Longer proofs are found in section 7

## 2 Definitions and Basic Concepts

A *directed cycle* is a sequence of edges $x \to \cdots \to x$. If a directed mixed graph $\mathcal{G}$ contains no directed cycles then it is *acyclic*. There may be two edges between a pair of vertices in an acyclic directed mixed graph, but multiple edges of the same type are not permitted; thus in this case one edge must be bi-directed ($x \leftrightarrow y$), otherwise there would be a directed cycle. The induced subgraph of $\mathcal{G}$ given by set $A$, denoted $\mathcal{G}_A$, consists of the subgraph of $\mathcal{G}$ with vertex set $A$, and those edges in $\mathcal{G}$ with both endpoints in $A$. We denote the subgraph of $\mathcal{G}$ made by removing all directed edges by $\mathcal{G}_{\leftrightarrow}$.

For a vertex $x$ in a mixed graph $\mathcal{G}$,

$$\mathrm{pa}_{\mathcal{G}}(x) \equiv \{v \mid v \to x \text{ in } \mathcal{G}\},$$
$$\mathrm{ch}_{\mathcal{G}}(x) \equiv \{v \mid v \leftarrow x \text{ in } \mathcal{G}\}$$
$$\text{and} \quad \mathrm{sp}_{\mathcal{G}}(x) \equiv \{v \mid v \leftrightarrow x \text{ in } \mathcal{G}\}$$

are the sets *parents*, *children* and *spouses*[1] of $x$ respectively.

$$\mathrm{an}_{\mathcal{G}}(x) \equiv \{v \mid v \to \cdots \to x \text{ in } \mathcal{G} \text{ or } v = x\},$$
$$\mathrm{de}_{\mathcal{G}}(x) \equiv \{v \mid v \leftarrow \cdots \leftarrow x \text{ in } \mathcal{G} \text{ or } v = x\}$$
$$\text{and} \quad \mathrm{dis}_{\mathcal{G}}(x) \equiv \{v \mid v \leftrightarrow \cdots \leftrightarrow x \text{ in } \mathcal{G} \text{ or } v = x\}.$$

are the set of *ancestors*, the set of *descendants* and the *district* of $x$ respectively. A district of $\mathcal{G}$ is a connected set in $\mathcal{G}_{\leftrightarrow}$. These definitions are applied disjunctively to sets of vertices, so that, for example,

$$\mathrm{pa}_{\mathcal{G}}(A) \equiv \bigcup_{x \in A} \mathrm{pa}_{\mathcal{G}}(x), \qquad \mathrm{sp}_{\mathcal{G}}(A) \equiv \bigcup_{x \in A} \mathrm{sp}_{\mathcal{G}}(x).$$

(Note that $\mathrm{sp}_{\mathcal{G}}(A) \cap A$ may be non-empty, and likewise for the other definitions.) We write $\mathrm{pa}_A$ for $\mathrm{pa}_{\mathcal{G}_A}$ and similarly for other definitions; sometimes we will omit the subscript altogether where context allows. Lastly let $\mathrm{barren}_{\mathcal{G}}(A) = \{x \mid \mathrm{de}_{\mathcal{G}}(x) \cap A = \{x\}\}$, which we refer to as the *barren* subset of $A$; a set $A$ is said to be barren if $\mathrm{barren}_{\mathcal{G}}(A) = A$.

---
[1] This usage differs from that of some authors who use the term 'spouse' to denote the other parents of the children of a vertex in a DAG.

For a mixed graph $\mathcal{G}$ with vertex set $V$ we consider collections of random variables $(X_v)_{v \in V}$ taking values in $\{0,1\}^{|V|}$. We use the usual shorthand notation: $v$ denotes a vertex and a random variable $X_v$, likewise $A$ denotes a vertex set and $X_A$. We write $i_v$, $\boldsymbol{i}_A$ for fixed states in $\{0,1\}$, $\{0,1\}^{|A|}$, respectively. For a sequence of independent and identically distributed observations $X_V^{(1)}, \ldots, X_V^{(n)}$, we write $\boldsymbol{n} = (n_{\boldsymbol{i}})_{\boldsymbol{i} \in \{0,1\}^{|V|}}$ for the vector of counts $n_{\boldsymbol{i}} = \sum_{j=1}^{n} \mathbf{1}_{\{X_V^{(j)} = \boldsymbol{i}\}}$.

### 2.1 m-Separation

A *path* between $x$ and $y$ in $\mathcal{G}$ is a sequence of edges $\langle \epsilon_1, \ldots, \epsilon_n \rangle$, such that there exists a sequence of vertices $\langle x \equiv w_1, \ldots, w_{n+1} \equiv y \rangle$, $(n \geq 0)$, where edge $\epsilon_i$ has endpoints $w_i, w_{i+1}$ (paths consisting of a single vertex are permitted). It is necessary to specify a path as a sequence of edges rather than vertices because the latter does not specify a unique path when there may be two edges between a given pair of vertices. A path of the form $x \to \cdots \to y$ is a *directed path* from $x$ to $y$.

A non-endpoint vertex $v$ on a path $\boldsymbol{\pi}$ is called a *collider* if the edges adjacent to it on $\boldsymbol{\pi}$ both have arrow heads pointing to $v$; i.e. it must be of the form $\to v \leftarrow$, $\leftrightarrow v \leftarrow$, $\to v \leftrightarrow$ or $\leftrightarrow v \leftrightarrow$; otherwise $v$ is a *non-collider*.

A path from $x$ to $y$ is said to be *m-connecting* given a set $Z$ if both

(i) no non-collider is in $Z$, and

(ii) every collider has a descendant in $Z$.

Otherwise it fails to m-connect. If there is no path m-connecting $x$ and $y$ given $Z$, then $x$ and $y$ are said to be *m-separated* given $Z$. Sets $X$ and $Y$ are said to be m-separated given $Z$ if for every pair $x \in X$ and $y \in Y$, $x$ and $y$ are m-separated given $Z$. Note that this definition includes the case $Z = \emptyset$, requiring only that a path has no colliders to be m-connecting. This is a natural extension of the d-separation criterion for DAGs; see for example Pearl (1988).

A probability measure $P$ on $\{0,1\}^{|V|}$ is said to satisfy the *global Markov property* for $\mathcal{G}$ if for every triple of disjoint sets $(X, Y, Z)$, where $Z$ may be empty, $X$ is m-separated from $Y$ given $Z$ implies $X \perp\!\!\!\perp Y | Z$ under $P$.

### 2.2 Heads and Tails

A non-empty set $H \subseteq V$ is a *head* in $\mathcal{G}$ if it is contained within a single district of $\mathcal{G}$ and is barren. We denote the collection of all heads by $\mathcal{H}(\mathcal{G})$. The *tail* associated with $H$ is defined as

$$\mathrm{tail}(H) \equiv \big(\mathrm{dis}_{\mathrm{an}(H)}(H) \setminus H\big) \cup \mathrm{pa}(\mathrm{dis}_{\mathrm{an}(H)}(H)).$$

Where context makes clear which head we are referring to, we will sometimes denote a tail by $T$. In $\mathcal{G}_1$, we have the following head and tail pairs:

| $H$ | $\{1\}$ | $\{2\}$ | $\{3\}$ | $\{2,3\}$ | $\{4\}$ | $\{3,4\}$ |
|---|---|---|---|---|---|---|
| $T$ | $\emptyset$ | $\{1\}$ | $\emptyset$ | $\{1\}$ | $\{2\}$ | $\{1,2\}$ |

In the case of a DAG, the heads are singleton sets $\{v\}$, and $\text{tail}(\{v\}) = \text{pa}_\mathcal{G}(v)$; in the case of a bi-directed graph, the heads are connected sets of vertices, and the tails are empty.

### 2.3 Partitions

In order to find the factorization of an ADMG, we must be able to partition any set of vertices into a collection of heads. This is done using the following functions. For an ADMG $\mathcal{G}$ and any subset of vertices $W$, define

$$\Phi_\mathcal{G}(\emptyset) \equiv \emptyset,$$
$$\Phi_\mathcal{G}(W) \equiv \left\{ H \,\middle|\, H = \bigcap_{x \in H} \text{barren}\left(\text{an}_\mathcal{G}\left(\text{dis}_W(x)\right)\right); H \neq \emptyset \right\},$$

$$\psi_\mathcal{G}(W) \equiv W \setminus \bigcup_{H \in \Phi_\mathcal{G}(W)} H,$$

$$\psi_\mathcal{G}^{(k)}(W) \equiv \underbrace{\psi_\mathcal{G}(\cdots \psi_\mathcal{G}(W) \cdots)}_{k\text{-times}}, \qquad \psi_\mathcal{G}^{(0)}(W) \equiv W.$$

Finally, we define the *partition induced by* $\mathcal{G}$ *on* $W$ to be:

$$[W]_\mathcal{G} \equiv \bigcup_{k \geq 0} \Phi_\mathcal{G}\left(\psi_\mathcal{G}^{(k)}(W)\right).$$

For a subset $W$, $\Phi$ takes the collection of barren nodes in each district; the partition is found by repeatedly removing these nodes and applying $\Phi$ again. For more details, including results on properties of $[\cdot]_\mathcal{G}$, see Richardson (2009).

## 3 Parametrization

A set $W \subseteq V$ is *ancestral* if $W = \text{an}_\mathcal{G}(W)$. It was shown by Richardson (2009) that a general probability distribution $F$ (not necessarily binary) obeys the global Markov property for $\mathcal{G}$ if and only if for every ancestral set $A$,

$$f_A(X_A) = \prod_{H \in [A]_\mathcal{G}} f_{H|T}(X_H \mid X_{\text{tail}(H)}),$$

where $f_A$ and $f_{H|T}$ are respectively marginal and conditional densities associated with $F$.

Returning to the binary case, we define the *generalized Möbius parameters* of $\mathcal{G}$ as the collection

$$\left\{ q_{H|\text{tail}(H)}^{i_{\text{tail}(H)}} \,\middle|\, H \in \mathcal{H}(\mathcal{G}),\, i_{\text{tail}(H)} \in \{0,1\}^{|\text{tail}(H)|} \right\}$$

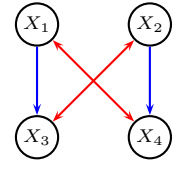

Figure 2: An ADMG with no vertex ordering under which the tail of each head precedes any vertex in the head.

where for $A, B \subseteq V$, $q_{A|B}^{i_B} = P(X_A = 0 \mid X_B = i_B)$. Richardson (2009) showed that a probability distribution $P$ obeying the global Markov property for an ADMG $\mathcal{G}$ is parametrized by this collection; since these are just conditional probabilities, everything is fully identified. In the case of a fully bi-directed graph, the generalized Möbius parameters become the ordinary Möbius parameters $q_A \equiv q_{A|\emptyset}^{i_\emptyset} = P(X_A = 0)$. In fact, under the global Markov property for $\mathcal{G}$,

$$P(X_V = i_V) = \sum_{C\,:\,O \subseteq C \subseteq V} (-1)^{|C \setminus O|} \prod_{H \in [C]_\mathcal{G}} q_{H|T}^{i_T} \quad (1)$$

where $O = \{v \in V : i_v = 0\}$.

**Lemma 1**
Suppose that $D_1 \cup D_2 \cup \cdots \cup D_k = V(\mathcal{G})$ and each pair $D_i$ and $D_j$, $i \neq j$ are disconnected in $\mathcal{G}_{\leftrightarrow}$. Further let $O_j = O \cap D_j$ for each $j$. Then

$$P(X_V = i_V) = \prod_{j=1}^k \sum_{C:O_j \subseteq C \subseteq D_j} (-1)^{|C \setminus O_j|} \prod_{H \in [C]_\mathcal{G}} q_{H|T}^{i_T}.$$

Thus we can factorize this parametrization into districts; note that this does *not* imply independence between districts, since the tail of a head in one district may contain vertices in another. Additionally, there is in general no partial order on heads such that each tail of a head is contained within the earlier heads. In figure 2, for example, the factorization is $f_{1234} = f_{23|1} \cdot f_{14|2}$. In the special case where $\mathcal{G}$ is a DAG, the districts are singleton sets of vertices $\{v\}$, each factor is of the form $P(X_v = i_v | X_{\text{pa}(v)} = i_{\text{pa}(v)})$, and the product is the familiar DAG factorization.

**Example 1**
For the graph $\mathcal{G}_1$ (figure 1), we have

$$\begin{aligned}
p_{1101} &= P(X_V = (1,1,0,1)^T) \\
&= q_3 - q_1\, q_3 - q_{23|1}^{(1)} - q_{34|12}^{(1,1)} + q_{23|1}^{(1)}\, q_1 \\
&\quad + q_{34|12}^{(1,1)}\, q_1 + q_{34|12}^{(1,1)}\, q_{2|1}^{(1)} - q_{34|12}^{(1,1)}\, q_{2|1}^{(1)}\, q_1 \\
&= (1 - q_1)(q_3 - q_{23|1}^{(1)} - q_{34|12}^{(1,1)} + q_{34|12}^{(1,1)}\, q_{2|1}^{(1)}) \\
&= P(X_1 = 1) \cdot P(X_{234} = (1,0,1)^T | X_1 = 1).
\end{aligned}$$

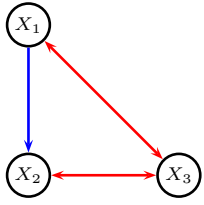

Figure 3: An ADMG, $\mathcal{G}_2$, used to illustrate construction of $M$ and $P$.

We refer to a product of the form

$$\prod_{H \in [C]_\mathcal{G}} q_{H|T}^{i_T}, \qquad (2)$$

as a *term*. From the definition of the partition $[\cdot]_\mathcal{G}$, it is clear that each term has exactly one factor whose head contains a given vertex $v \in C$; hence the expression for $p_{\boldsymbol{i}}$ is a multi-linear polynomial. Letting $\boldsymbol{q} = (q_{H|T}^{i_T} | H \in \mathcal{H}(\mathcal{G}))$, we show below that the expression (1) can be written in the form

$$\boldsymbol{p}(\boldsymbol{q}) = M \exp(P \log \boldsymbol{q}) \qquad (3)$$

for matrices $M$ and $P$, where the operations exp and log are taken pointwise over vectors. This is similar to the form of generalized log-linear models considered by Lang et al. (1999).

### 3.1 Construction of $M$ and $P$

We restrict our attention to graphs containing only one district. Then each $p_{\boldsymbol{i}}$ is simply a sum of terms of the form (2) (up to sign) characterized by $C$ and the tail states $\boldsymbol{i}_T$. We define a matrix $M$ whose rows correspond to the possible states $\boldsymbol{i}$, and whose columns correspond to possible terms of the form (2). Let $M$ have $(j,k)$th entry $\pm 1$ if the term associated with column $k$ appears with that coefficient in the expression for the probability associated with row $j$; otherwise the entry is 0. For example, in the graph $\mathcal{G}_2$ in figure 3,

$$p_{101} = q_{2|1}^{(1)} - q_{2|1}^{(1)} q_1 - q_{23|1}^{(1)} + q_{23|1}^{(1)} q_1.$$

Then the row of $M$ associated with the state $(1,0,1)^T$ contains entries

$$\begin{pmatrix} \emptyset & \{1\} & \{2\}_{i_1=0} & \{2\}_{i_1=1} & \{1,2\}_{i_1=0} & \{1,2\}_{i_1=1} & \{3\} \\ 0 & 0 & 0 & +1 & 0 & -1 & 0 \end{pmatrix}$$

$$\begin{pmatrix} \{1,3\} & \{2,3\}_{i_1=0} & \{2,3\}_{i_1=1} & \{1,2,3\}_{i_1=0} & \{1,2,3\}_{i_1=1} \\ 0 & 0 & -1 & 0 & +1 \end{pmatrix}.$$

We create a second matrix $P$ which contains a row for each term of the form (2), and a column for each element of $\boldsymbol{q}$; it will be used to map the generalized Möbius parameters to the terms. The $(j,k)$th entry in $P$ is 1 if the term associated with row $j$ contains the parameter associated with column $k$ as a factor. Thus in $\mathcal{G}_2$, for $C = \{1,2\}$ and $i_1 = 1$, the associated product is $q_{2|1}^{(1)} q_1$, and the associated row of $P$ contains the entries

$$\begin{pmatrix} q_1 & q_{2|1}^{(0)} & q_{2|1}^{(1)} & q_3 & q_{13} & q_{23|1}^{(0)} & q_{23|1}^{(1)} \\ 1 & 0 & 1 & 0 & 0 & 0 & 0 \end{pmatrix},$$

where the parameters are shown above their respective columns.

Then it is clear that the operation $\exp(P \log \boldsymbol{q})$ maps the vector of parameters $\boldsymbol{q}$ to a vector containing all possible terms. The full matrix $P$ for $\mathcal{G}_2$ is as follows:

|  | $q_1$ | $q_{2|1}^{(0)}$ | $q_{2|1}^{(1)}$ | $q_3$ | $q_{13}$ | $q_{23|1}^{(0)}$ | $q_{23|1}^{(1)}$ |
|---|---|---|---|---|---|---|---|
| $\emptyset$ | 0 | 0 | 0 | 0 | 0 | 0 | 0 |
| $\{1\}$ | 1 | 0 | 0 | 0 | 0 | 0 | 0 |
| $\{2\}, i_1=0$ | 0 | 1 | 0 | 0 | 0 | 0 | 0 |
| $\{2\}, i_1=1$ | 0 | 0 | 1 | 0 | 0 | 0 | 0 |
| $\{1,2\}, i_1=0$ | 1 | 1 | 0 | 0 | 0 | 0 | 0 |
| $\{1,2\}, i_1=1$ | 1 | 0 | 1 | 0 | 0 | 0 | 0 |
| $\{3\}$ | 0 | 0 | 0 | 1 | 0 | 0 | 0 |
| $\{1,3\}$ | 0 | 0 | 0 | 0 | 1 | 0 | 0 |
| $\{2,3\}, i_1=0$ | 0 | 0 | 0 | 0 | 0 | 1 | 0 |
| $\{2,3\}, i_1=1$ | 0 | 0 | 0 | 0 | 0 | 0 | 1 |
| $\{1,2,3\}, i_1=0$ | 1 | 0 | 0 | 0 | 0 | 1 | 0 |
| $\{1,2,3\}, i_1=1$ | 1 | 0 | 0 | 0 | 0 | 0 | 1 |

For a graph with multiple districts, it is most efficient to construct a pair $(M^j, P^j)$ for each district $j$, and then

$$\boldsymbol{p}(\boldsymbol{q}) = \prod_j M^j \exp(P^j \log \boldsymbol{q}^j)$$

using the result of lemma 1; here $\boldsymbol{q}^j$ is a vector of parameters whose heads are in the district $j$.

The number of columns in $M$ is $\sum_{H \in \mathcal{H}} 2^{|\operatorname{tail}(H)|}$, but there are at most $|\mathcal{H}|$ entries in each row. Hence $M$ becomes large quickly for large districts, but is also sparse, especially if tail sizes are large. Similar comments apply to $P$.

### 3.2 Constraint Matrix

The likelihood of the model associated with a graph $\mathcal{G}$ has the form

$$l(\boldsymbol{q}) = \sum_{\boldsymbol{i} \in \{0,1\}^{|V|}} n_{\boldsymbol{i}} \log p_{\boldsymbol{i}}(\boldsymbol{q}),$$

where $p(q)$ is the smooth function defined by (3), and $p_i(q)$ is the element of $p$ corresponding to the state $i$. For the purposes of the maximum likelihood estimation algorithm, we will need to ensure that we remain within the valid parameter space. Let $\mathcal{P}_\mathcal{G}$ be the collection of all vectors of probabilities $p$ in the simplex $\Delta_{2^{|V|}-1}$ which satisfy the global Markov property for $\mathcal{G}$; let $\mathcal{Q}_\mathcal{G}$ be the image of $\mathcal{P}_\mathcal{G}$ under the map $p^{-1} = q$ which takes probabilities to generalized Möbius parameters.

The definition of the generalized Möbius parameters as conditional probabilities makes it clear that this inverse $q : \mathcal{P}_\mathcal{G} \to \mathcal{Q}_\mathcal{G}$ is also smooth if all $p_i > 0$. The form of the derivative is given in lemma 3 in section 7.

It is not easy to characterize $\mathcal{Q}_\mathcal{G}$ directly; clearly $q \geq 0$, but other constraints such as $q_{23|1}^{(0)} \leq q_{2|1}^{(0)}$ may be more subtle. We approach this by considering one vertex $v$ at a time, updating the parameters whose heads contain $v$: $\theta^v \equiv (q_{H|T}^{i_T} \mid v \in H \in \mathcal{H}(\mathcal{G}))$; the rest are held fixed. When updating $\theta^v$ we need to impose constraints in order to ensure that the associated complete parameter vector remains in $\mathcal{Q}_\mathcal{G}$.

Since each term in the map $p_i(q)$ contains at most one factor with $v$ in its head, we have that $p$ is a linear function of $\theta^v$; i.e. $p = A^v \theta^v - b^v$ for some matrix $A^v$ and vector $b^v$. We need to ensure that $p_i \geq 0$ for each $i$, so the constraints amount to

$$A^v \theta^v \geq b^v, \qquad (4)$$

where the inequality is understood to act pointwise on vectors. Note that $\sum_i p_i(q) = 1$ is guaranteed by the form of $p(q)$.

For graphs with multiple districts, the value of $A^v$ where $v$ is in a district $D$ will not depend upon parameters associated with other districts.

## 4  Maximum Likelihood Fitting Algorithm

The basis of our algorithm is a block co-ordinate updating scheme with gradient ascent. For simplicity we will restrict attention to the case where all the counts $n = (n_i)_{i \in \{0,1\}^{|V|}}$ are strictly positive. This will ensure that $p$ is strictly positive, and the possibility of optima on the boundary need not be taken into account. In the case of zero counts, which will be considered in future work, the partial likelihood function that is considered below is still concave but need no longer be strictly concave.

At each step we will increase the likelihood by changing the parameters associated with each vertex, considering each vertex in turn; specifically we will update only the generalized Möbius parameters which contain $v$ in their head. Then the partial likelihood has the form

$$l(\theta^v) = \sum_i n_i \log p_i^v(\theta^v)$$

where $p_i^v$ are purely linear functions in $\theta^v$. Then the likelihood is strictly concave in $\theta^v$, and hence this can be solved using a gradient ascent approach, subject to the linear constraints $A^v \theta^v \geq b^v$. A feasible starting value is easily found using, for example, full independence.

**Algorithm 1**

Cycle through each vertex $v \in V$ performing the following steps:

**Step 1.** Construct the constraint matrix $A^v$.

**Step 2.** Solve the non-linear program

$$\begin{aligned}
\text{maximize} \quad & l(\theta^v) = \sum_i n_i \log p_i^v(\theta^v) \\
\text{subject to} \quad & A^v \theta^v \geq b^v.
\end{aligned}$$

Stop when a complete cycle of $V$ results in a sufficiently small increase in the likelihood.  □

The problem in step 2 has a unique solution $\theta^v$, and is easy to solve using a gradient ascent method; a line search using the Armijo rule ensures convergence. See Bertsekas (1999) for examples. The likelihood is guaranteed not to decrease at each step, and if the algorithm cycles through all vertices $v$ without moving, we are guaranteed to have reached a (possibly local) maximum.

In the event that we move to a point $q \notin \mathcal{Q}$ with $p(q) \in \Delta_{2^{|V|}-1}$, we simply map to the point $q' = q(p(q))$, which is guaranteed to be in $\mathcal{Q}_\mathcal{G}$. Thus it is sufficient to ensure that $p(q) \in \Delta_{2^{|V|}-1}$. For graphs with more than one district, we can apply Algorithm 1 to each district, possibly in parallel.

A 'black box' fitting algorithm could also be used to find ML estimates; however our approach gives more clarity to the parameterization and fitting problem, and we anticipate that it will be useful for implementing extensions to these models.

**Standard Errors**

Since this is a curved exponential family, standard errors can be obtained from the Fisher information matrix, $I(q)$ (Johansen, 1979). Let $p^\dagger = p(q^\dagger)$ be the 'true' probability distribution of $X_V$, and $\hat{p} = p(\hat{q})$ the maximum likelihood estimate. Define the augmented

likelihood $l_\lambda$ for the sample by

$$l_\lambda(\boldsymbol{p}) = \sum_i n_i \log p_i + \lambda \left(1 - \sum_i p_i\right)$$

and note that $\nabla_{\boldsymbol{q}} l_\lambda = \nabla_{\boldsymbol{q}} l$ since $1 - \sum_i p_i(\boldsymbol{q}) = 0$ for all $\boldsymbol{q}$. Then we have

$$\nabla_{\boldsymbol{q}} l_\lambda(\boldsymbol{p}(\boldsymbol{q})) = \frac{\partial \boldsymbol{p}}{\partial \boldsymbol{q}} \cdot \nabla_{\boldsymbol{p}} l_\lambda(\boldsymbol{p}),$$

where

$$\frac{\partial l}{\partial p_i} = n_i p_i^{-1} - \lambda.$$

Then choosing $\lambda = n$ gives $\mathbb{E}_{\boldsymbol{p}^\dagger}\left[\nabla_{\boldsymbol{p}} l_\lambda(\boldsymbol{p}^\dagger)\right] = \boldsymbol{0}$, and

$$n^{-1} \mathbb{E}_{\boldsymbol{p}^\dagger}\left[(\nabla_{\boldsymbol{p}} l_\lambda)(\nabla_{\boldsymbol{p}} l_\lambda)^T\right] = \operatorname{diag} 1/\boldsymbol{p}^\dagger - \boldsymbol{1}\boldsymbol{1}^T,$$

where $(1/\boldsymbol{p})_i = 1/p_i$. Thus

$$I(\boldsymbol{q}^\dagger) = n^{-1} \mathbb{E}_{\boldsymbol{p}(\boldsymbol{q}^\dagger)}\left[(\nabla_{\boldsymbol{q}} l)(\nabla_{\boldsymbol{q}} l)^T\right]$$
$$= \left(\left.\frac{\partial \boldsymbol{p}}{\partial \boldsymbol{q}}\right|_{\boldsymbol{q}=\boldsymbol{q}^\dagger}\right)(\operatorname{diag} 1/\boldsymbol{p}^\dagger - \boldsymbol{1}\boldsymbol{1}^T)\left(\left.\frac{\partial \boldsymbol{p}}{\partial \boldsymbol{q}}\right|_{\boldsymbol{q}=\boldsymbol{q}^\dagger}\right)^T.$$

An expression for $\frac{\partial \boldsymbol{p}}{\partial \boldsymbol{q}}$ is given in lemma 3. Then $\sqrt{n}(\hat{\boldsymbol{q}} - \boldsymbol{q}^\dagger) \xrightarrow{\mathcal{D}} \mathcal{N}(\boldsymbol{0}, I(\boldsymbol{q}^\dagger)^{-1})$ and we approximate the standard error of $q_j$ by $\sqrt{[I(\hat{\boldsymbol{q}})^{-1}]_{jj}}$.

## 5 Examples

As an illustration of our method, we fit ADMGs to data from the US General Social Survey between 1975 and 1994. The data have seven outcomes which have been dichotomized; briefly these are:

**Trust:** can most people be trusted?
**Helpful:** are people generally helpful?
**ConLegis:** do you have confidence in congress?
**ConClerg:** ...in organized religion?
**ConBus:** ...in major companies?
**MemUnion:** are you a member of a labor union?
**MemChurch:** ...of a church?

For further details and the full dataset, see Drton and Richardson (2008a).

We use a simple stepwise selection procedure based on BIC, starting with $\mathcal{G}^{(0)}$, the graph with no edges. Given the current best fitting graph $\mathcal{G}^{(t)}$, all graphs which differ from $\mathcal{G}$ only by addition or removal of one edge are considered. If none of the considered graphs has a lower BIC when fitted to the data than $\mathcal{G}^{(t)}$, then the procedure stops; otherwise the graph with the lowest BIC becomes $\mathcal{G}^{(t+1)}$. There are some subtleties, since there may be several Markov equivalent graphs to choose between: for example, when picking the first edge it makes no difference whether we add a directed or bi-directed edge. In future work we will consider equivalence classes of graphs which impose the same independence structure.

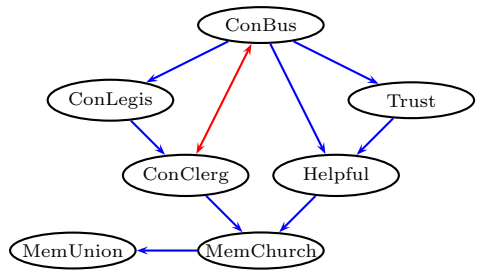

Figure 4: Best fitting graph by BIC, from stepwise procedure.

The algorithm selects the graph shown in figure 4; the graph is equivalent to a DAG, simply by replacing ConBus $\leftrightarrow$ ConClerg with ConBus $\rightarrow$ ConClerg. The selected graph is quite different to that of Drton and Richardson (2008a), who used likelihood ratios with backwards selection to find their model. The deviance is 180.8 on 108 degrees of freedom, which gives evidence to reject the model in favor of the saturated model ($p < 10^{-4}$).

A similar procedure using AIC produced the graph in figure 5; this has more parameters, and is closer to the graph of Drton and Richardson (2008a). The deviance is 76.7 on 76 degrees of freedom, suggesting that the model is a good fit for the data ($p = 0.456$). Note that although there is no edge between ConBus and MemChurch, nor between ConClerg and Helpful, there is no set of vertices which make either of these pairs independent under the global Markov property. This is because ConBus $\rightarrow$ Trust $\leftrightarrow$ MemChurch forms an inducing path (Richardson and Spirtes, 2003); similarly for ConClerg $\leftrightarrow$ ConBus $\leftrightarrow$ Helpful. Indeed, this graph is also equivalent to a DAG.

The graph in figure 6 was fitted to data on recidivism (Fienberg, 1980, p. 93) using the stepwise procedure for AIC; this graph is not equivalent to a DAG.

## 6 Computation

We implemented the algorithm in the statistical computing package R (R Development Core Team, 2004). The program was generally able to fit graphs with largest district size at most 5 in under a minute running on a desktop computer with a Pentium 4 3.4 GHz processor and 2 GB of RAM. We fit graphs of the forms shown in figure 8, for increasing values of $k$.

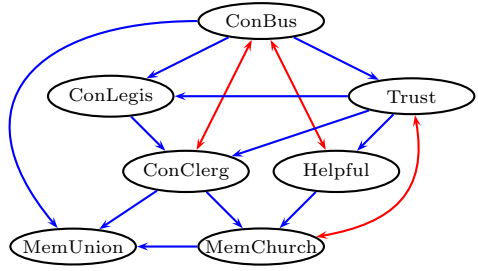

Figure 5: Best fitting graph by AIC, from stepwise procedure.

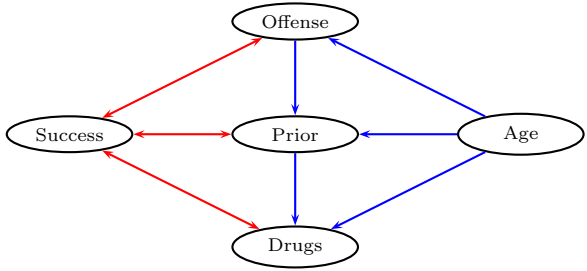

Figure 6: Best fitting graph for recidivism data by AIC, from stepwise procedure.

An average of the elapsed time in seconds over the 10 runs is plotted in figure 7. Note that the graph with fixed district sizes is fitted much more quickly, which is not surprising since it limits the size of the matrices $M$ and $P$. This suggests that, given a set of Markov equivalent graphs, it may be most efficient to use the one with the smallest districts, and/or fewest arrowheads (Ali et al., 2005; Drton and Richardson, 2008b). In the easiest case where all the districts have size one, we have a DAG, which is fitted after only one cycle through the vertices.

## 7  Proofs and Technical Results

**Proof of Lemma 1.** We prove the case $k = 2$, from which the full result follows trivially by induction.

The operation $[C]_{\mathcal{G}}$ partitions into sets which are $\mathcal{G}_{\leftrightarrow}$-connected, thus

$$P(X_V = \boldsymbol{i}_V)$$
$$= \sum_{C:O\subseteq C\subseteq V} (-1)^{|C\setminus O|} \left[ \prod_{H_1 \in [C\cap D_1]_{\mathcal{G}}} q^{i_{T_1}}_{H_1|T_1} \prod_{H_2 \in [C\cap D_2]_{\mathcal{G}}} q^{i_{T_2}}_{H_2|T_2} \right].$$

Also,

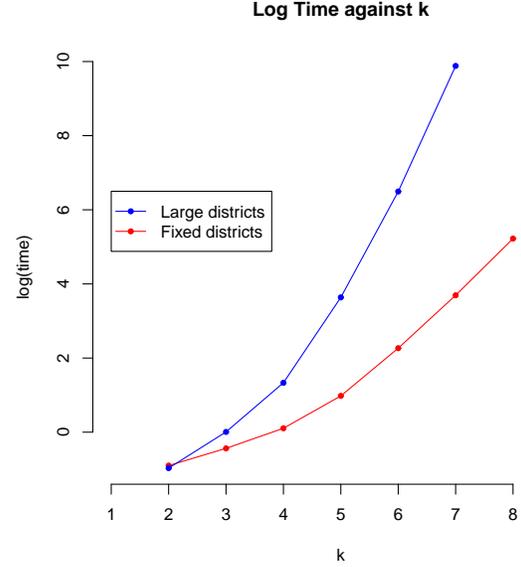

Figure 7: Log time in seconds taken to fit the graphs in figure 8 for various values of $k$. Average of 10 runs except for $k = 7$ for large districts, average of 2 runs.

$$\{C : O \subseteq C \subseteq V\}$$
$$= \{C_1 \cup C_2 : O_1 \subseteq C_1 \subseteq D_1,\ O_2 \subseteq C_2 \subseteq D_2\}$$

and $C \cap D_i = C_i$, so

$$P(X_V = \boldsymbol{i}_V)$$
$$= \sum_{\substack{C_1:O_1\subseteq C_1\subseteq D_1 \\ C_2:O_2\subseteq C_2\subseteq D_2}} (-1)^{|(C_1\cup C_2)\setminus O|} \left[ \prod_{H_1 \in [C_1]_{\mathcal{G}}} q^{i_{T_1}}_{H_1|T_1} \prod_{H_2 \in [C_2]_{\mathcal{G}}} q^{i_{T_2}}_{H_2|T_2} \right].$$

Noting that
$$|(C_1 \cup C_2) \setminus O| = |(C_1 \setminus O_1) \cup (C_2 \setminus O_2)|$$
$$= |C_1 \setminus O_1| + |C_2 \setminus O_2|,$$

we have

$$P(X_V = \boldsymbol{i}_V)$$
$$= \left\{ \sum_{C_1:O_1\subseteq C_1\subseteq D_1} (-1)^{|C_1\setminus O_1|} \prod_{H_1 \in [C_1]_{\mathcal{G}}} q^{i_{T_1}}_{H_1|T_1} \right\}$$
$$\times \left\{ \sum_{C_2:O_2\subseteq C_2\subseteq D_2} (-1)^{|C_2\setminus O_2|} \prod_{H_2 \in [C_2]_{\mathcal{G}}} q^{i_{T_2}}_{H_2|T_2} \right\}.$$

□

**Lemma 2**
Let $f_1, \ldots, f_k : \mathbb{R}^m \to \mathbb{R}$ and $g : \mathbb{R} \to \mathbb{R}$ be differentiable real-valued functions. Then writing

$$g(\boldsymbol{f}(\boldsymbol{x})) = \begin{pmatrix} g(f_1(\boldsymbol{x})) \\ \vdots \\ g(f_k(\boldsymbol{x})) \end{pmatrix}$$

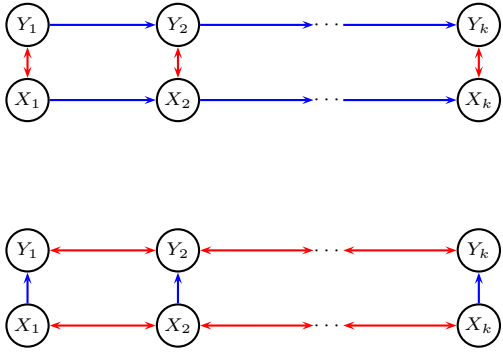

Figure 8: Expandable graphs with (a) fixed districts (top), (b) large districts (bottom).

we have

$$\nabla_{\boldsymbol{x}} g(\boldsymbol{f}(\boldsymbol{x})) = \operatorname{diag} g'(\boldsymbol{f}(\boldsymbol{x})) \cdot \frac{\partial \boldsymbol{f}(\boldsymbol{x})}{\partial \boldsymbol{x}}$$

where $g'$ is understood to act pointwise on vectors in the same way as $g$.

**Proof**

$$\begin{aligned}
[\nabla_{\boldsymbol{x}} g(\boldsymbol{f}(\boldsymbol{x}))]_{ij} &= \frac{\partial g(f_i(\boldsymbol{x}))}{\partial x_j} \\
&= \frac{\partial g(f_i(\boldsymbol{x}))}{\partial f_i(\boldsymbol{x})} \frac{\partial f_i(\boldsymbol{x})}{\partial x_j} \\
&= g'(f_i(\boldsymbol{x})) \cdot \left[\frac{\partial \boldsymbol{f}(\boldsymbol{x})}{\partial \boldsymbol{x}}\right]_{ij}
\end{aligned}$$

so we simply have $\frac{\partial \boldsymbol{f}(\boldsymbol{x})}{\partial \boldsymbol{x}}$ with the $i$th row scaled by a factor $g'(\boldsymbol{f}_i(\boldsymbol{x}))$; this is achieved using the diagonal matrix. $\square$

The following result follows directly from lemma 2.

**Lemma 3**

$$\frac{\partial \boldsymbol{p}}{\partial \boldsymbol{q}} = M \operatorname{diag}\left[\exp(P \log \boldsymbol{q})\right] P \frac{1}{\boldsymbol{q}}$$

where $\frac{1}{\boldsymbol{q}}$ is a vector with $j$th element $1/q_j$.

**Acknowledgements**

This research was supported by the U.S. National Science Foundation grant CNS-0855230 and U.S. National Institutes of Health grant R01 AI032475.